\documentclass[aps,prl,showpacs,showkeys,twocolumn]{revtex4}

\usepackage{graphicx}
\usepackage{latexsym}

\begin{document}

\title{Fractal analysis on a closed classical 
hard-wall billiard using a simplified box-counting algorithm}

\author{Suhan Ree}
\email{suhan@kongju.ac.kr}
\affiliation{Department of Industrial Information,
Kongju National University, Yesan-Up, Yesan-Gun, 
Chungnam, 340-802, South Korea}

\date{\today}

\begin{abstract}
We perform fractal analysis on a closed classical
hard-wall billiard, the circular billiard with a straight cut, assuming
there are two openings on the boundary.
We use a two-dimensional set of initial conditions
that produce all possible trajectories of a particle injected from one opening,
and numerically compute the fractal dimension of singular points of a function
that maps an initial condition to
the number of collisions with the wall before the exit.
We introduce a simplified box-counting algorithm, which uses points from
a rectangular grid inside the two-dimensional set of the initial conditions, to
simplify the calculation,
and observe the classical chaotic properties 
while varying the parameters of the billiard.
\end{abstract}

\pacs{05.45.Df, 05.45.Pq, 73.23.Ad}
\keywords{Chaos, Billiard, Classical scattering, Fractal}

\maketitle

Chaotic systems have recently attracted many researchers from
various fields,
partly because the fast development of computer hardwares
has enabled us to solve equations 
almost insolvable in the past\cite{ott,reichl}.
For physicists, the two-dimensional (2D)  billiard system 
has been a popular subject
for studying the dynamics of the Hamiltonian chaotic systems.
Classically, the dynamics of the 2D billiard system shows three 
distinct types of behaviors.
The system is either integrable (\emph{regular} behavior) or 
non-integrable 
(either \emph{soft chaos}, characterized by mixed phase spaces 
that have 
both regular and chaotic regions, or \emph{hard chaos}, 
characterized by ergodicity and mixing)\cite{gutzwiller}. 

Here we use a closed 2D hard-wall billiard
with two openings on the boundary, and perform fractal analysis using the 
classical dynamics
while varying the shape of the bliiard and the size of the openings.
To calculate the fractal dimension,
we use a set of initial conditions that will produce trajectories
of a particle injected from one opening,
and calculate certain values for each trajectory, such as 
the \emph{exit opening} (an opening from which 
the particle exits), the number of collisions with the wall 
before the exit,
the dwell time, and so on.
When the system is chaotic, singular points of this kind of response 
functions form a fractal set\cite{eckhardt,bleher,ott2}.
In this Letter, we will calculate the fractal dimension of 
these singular points of the function that maps an initial condition
to the number of collisions,
by introducing a simplified box-counting algorithm.

\begin{figure}
\includegraphics[scale=0.7]{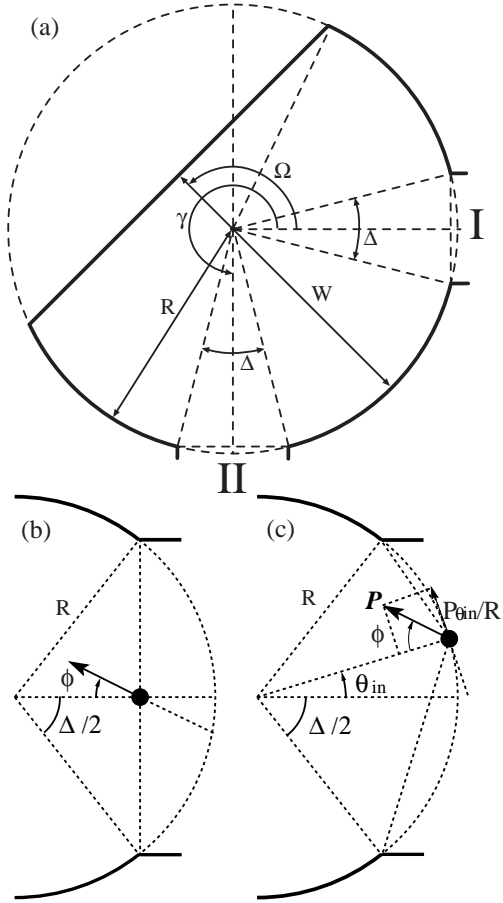}
\caption{\label{geometry}
(a) Geometry of the cut-circle billiard. There are
two openings, I and II, and a straight cut. 
The size of the cut is given by the width $W$ (here we use $w\equiv W/R$
as the relative size of the cut), and 
the size of two openings is $\Delta$.
The positions of the cut and the opening II, $\Omega$ and $\gamma$,
are measured from the position of the opening I. 
(b) When a particle is injected 
from the central point of the opening I with an angle $\phi$, the set of possible
initial conditions is one dimensional $(-\pi/2 < \phi < \pi/2)$.
(c) To fully represent all significant initial conditions, 
a particle is injected from a point on the circular boundary of the 
opening I with the angular momentum $P_{\theta \rm in}$. This set, whose element is
represented by
a pair of values, $(\theta_{\rm in},P_{\theta \rm in})$, is 
a closed two-dimensional subspace of the four-dimensional phase space.
}
\end{figure}

The billiard we have chosen is the circular billiard with a straight cut, 
or the ``cut-circle'' billiard (see Fig.\ \ref{geometry}).
There are five parameters: (1) the width $W$ measured in the
direction perpendicular to the cut, (2) the radius $R$, (3) the angular
width $\Delta$ of the openings, which represents the opening size, 
(4) the orientation angle ${\Omega}$ of the cut relative
to the first opening, and (5) the position of the second 
opening relative to the
first opening as measured by the angle $\gamma$. 
We scale the width $W$ by $R$, so
$w\equiv W/R$, thereby reducing the number of independent 
parameters to four:
$w$, $\Delta$, $\Omega$, and $\gamma$. 
For all subsequent discussions, we set $\Omega=135^\circ$ and $\gamma=270^\circ$.
For the closed cut-circle billiard (when $\Delta=0$), it has been proved in Ref.\ 
\onlinecite{bunimovich} that the phase 
space is mixed (soft chaos) when $0<w<1$, and that the phase space is fully chaotic
(hard chaos) when $1<w<2$.  
And it is integrable when $w=1$ and 2.
Consequently this billiard has a single shape parameter $w$
that makes the billiard exhibit all three types of chaotic behaviors.

In Ref.\ \onlinecite{ree7}, a one-dimensional (1D) set of initial conditions
of the cut-circle billiard
was used to calculate fractal dimensions while varying the opening size $\Delta$ and the
shape parameter $w$ [see Fig.\ \ref{geometry}(b)].
The particle is injected with an incident angle $\phi$ ($-\pi/2<\phi<\pi/2$).
In Ref.\ \onlinecite{legrand2}, authors used a 1D set of initial conditions 
for the \emph{stadium}, 
and calculated the recurrence time for each trajectory to obtain the fractal dimension.
These 1D sets, however, are subsets of the set of all significant initial conditions
that represent all possible trajectories; 
in our billiard, a 2D subspace in the four-dimensional (4D) phase space 
is a minimal set that will produce all possible trajectories originating 
from one opening [see Fig.\ \ref{geometry}(c)].
(A 2D set of initial conditions were also used in Ref.\ \onlinecite{bleher}
for the Sinai billiard.)
There are two reasons that make this reduction possible.
First, the value of the energy does not change the trajectory of the particle. 
This is due to the flat
potential inside the billiard with infinitely hard boundaries. 
Second, the distance from
the center of the billiard to the initial location of the particle can be
fixed at $R$. 

\begin{figure}
\includegraphics[scale=0.45]{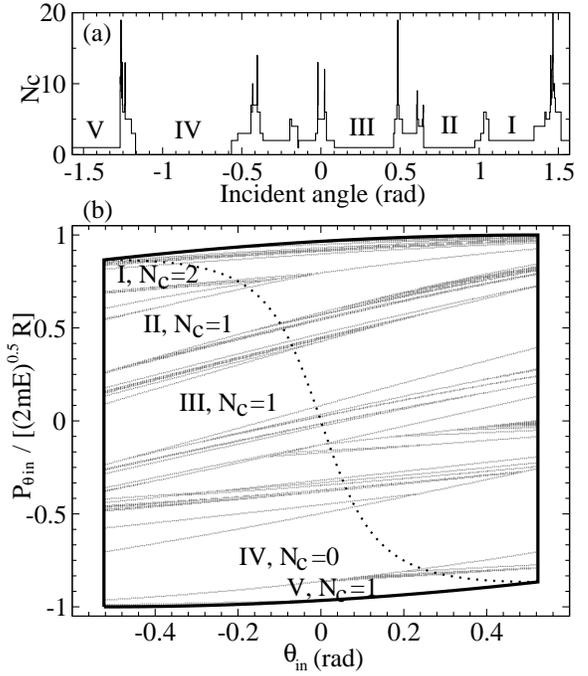}
\caption{\label{ncols}
Calculations of the number of collisions when $\Omega=135^\circ$,
$\gamma=270^\circ$, $w=1.5$, and $\Delta=60^\circ$.
(a) Graph of the incident angle  $\phi$ vs
the number of collisions $N_C$ 
when the 
particle is injected as in Fig.\ \ref{geometry}(b).
(b) When the particle is injected as in Fig.\ \ref{geometry}(c), all significant
initial conditions lie inside the closed two-dimensional space bounded 
by the bold line in the graph. In the graph, only singular points are shown.
The dotted line represents the
one-dimensional subset of initial conditions used in (a). 
Five geometrical channels with low $N_C$'s are also shown.
}
\end{figure}

With these two kinds of sets, we numerically find a function that maps an initial
condition to the number of collisions before the exit.
(In numerical computations, all possible initial conditions cannot be used;
only a finite and discrete subset can be considered.)
In Fig.\ \ref{ncols}, we show how the number of collisions changes for these
sets of initial conditions when $w=1.5$ and $\Delta=60^\circ$.
In Fig.\ \ref{ncols}(a), the graph of $\phi$ vs the number of collisions is 
shown when the 1D subset of initial conditions as in Fig.\ \ref{geometry}(b)
is used (see Ref.\ \onlinecite{ree7} for more results).
In Fig.\ \ref{ncols}(b), the graph shows the full 2D set of initial conditions,
$(\theta_{\rm in},P_{\theta\rm in})$, bounded by two equations,
\begin{equation}
-\frac{\Delta}{2} \le  \theta{\rm in}  \le \frac{\Delta}{2},
\end{equation}
\begin{equation}
-\cos\left(\frac{\Delta}{4}+\frac{\theta_{\rm in}}{2}\right) \le 
\frac{P_{\theta\rm in}}{\sqrt{2mE}R} \le \cos\left(\frac{\Delta}{4}-
\frac{\theta_{\rm in}}{2}\right),
\end{equation}
which are represented by the bold line.
Instead of the number of collisions,
the graph shows singular points that constitute singular boundaries
at which the number of collisions changes.
We can observe that singular points have infinitely fine structures.
The number of collisions are shown only for five 
\emph{geometrical channels}\cite{roukes,luna} in this example.
The dotted line is the 1D subset used in Fig.\ \ref{ncols}(a).

Now we calculate the fractal dimension $d_f$ of
a set of singular points in this 2D set using a simplified box-counting algorithm.
As a generalization of the box-counting algorithm used in Ref.\ \onlinecite{ree7}
for the 1D set, 
we use a uniform 2D rectangular grid. 
Then there are uniform rectangular boxes, and each of them is  
represented by four points on the grid.
For all of these uniformly distributed grid points inside the 2D set, 
we numerically find
the number of collisions before the exit,
and for each box, we compare the number of collisions at four corners.
With an assumption that a box does not contain a singular point
when four values are all equal, we can count the number of boxes
containing singularities out of all boxes inside the 2D set.
The number of all boxes inside the set is $N_P$, and 
the number of boxes that contain any singular point, based on the 
above assumption, is $N_S$.
Then the fractal dimension $d_f$ is defined by 
\begin{equation}
\label{df}
d_f\equiv\lim_{N_P\rightarrow\infty}\frac{\log_{10}{N_S}}{\log_{10}{N_P}^{1/2}},
\end{equation}
which is the slope of the graph of $0.5\log_{10}{N_P}$ vs $\log_{10}{N_S}$.
In numerical calculations, we find $N_S$ for several $N_P$ values,
and then use the ordinary least-square fit 
to find the slope using the points in the graph. 

\begin{figure}
\includegraphics[scale=0.7]{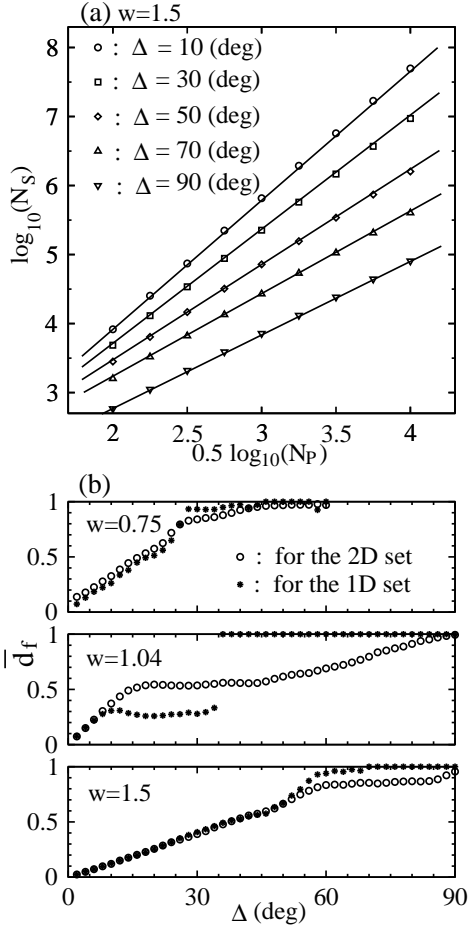}
\caption{\label{angles}
(a) When $w=1.5$, 
the number of boxes containing the singular points, $N_S$, is counted
as the number of boxes, $N_P$, increases, 
for five different $\Delta$ values.
Then $d_f$ can be found from the slopes of these log-log graphs.
(b) Graphs of $\bar{d_f}$ vs the opening size $\Delta$ 
for $w=0.75$, $1.04$, and $1.5$.
Here we compare the values  of $\bar{d_f}$ for the 2D set ($\circ$) and 
for the 1D set ($*$). 
}
\end{figure}

In Fig.\ \ref{angles}(a), the log-log graph is shown for several $\Delta$ values
when $w=1.5$. (For all calculations, $N_P$ up to $10^8$ will be used.)
Unlike the calculations using the 1D set, $d_f$ will be in the range of 
$1\le d_f \le 2$.
We can define $\bar{d_f}$, which better represents the fractality,
\begin{equation}
\bar{d_f}\equiv \left\{ \begin{array}{rl}
2-d_f~&\mbox{(for the 2D set)}\\
1-d_f~&\mbox{(for the 1D set)}
\end{array} \right..
\end{equation}
In Fig.\ \ref{angles}(b), $\bar{d_f}$'s both for the 2D set and for the 1D set 
are compared for three different $w$ values while varying $\Delta$.
Like the curves for the 1D set,
not all curves monotonically increase
as $\Delta$ increases, even though it is not as significant.
The reason for this phenomenon is that there are more possible trajectories
when the opening gets bigger, and it also explains why
curves don't match very well for large $\Delta$.
We also observe that the cases with hard chaos ($w=1.04$, and 1.5) do not behave 
predictably for large $\Delta$, which will be more clearly seen in the following
graph.
When the original shape is changed sufficiently by the presence of the
openings, the ergodicity of the billiard 
is no longer an important factor; the relative locations of the 
openings start to
have more effect on the fractal dimension as in the cases of soft chaos.

\begin{figure}
\includegraphics[scale=0.45]{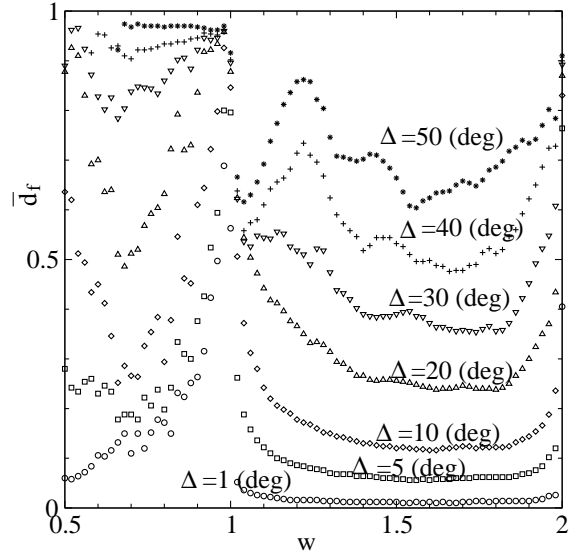}
\caption{\label{widths} 
Graphs of $w$ vs $\bar{d_f}$ for seven different 
opening sizes: $\Delta=1^\circ$ ($\circ$), $\Delta=5^\circ$ ($\Box$), 
$\Delta=10^\circ$ ($\Diamond$),
$\Delta=20^\circ$ ($\bigtriangleup$), $\Delta=30^\circ$ ($\bigtriangledown$), 
$\Delta=40^\circ$ ($+$), and
$\Delta=50^\circ$ ($*$), for the 2D set.
When $\Delta$ is small enough ($\Delta\le20^\circ$ in the graph), 
the behavior of the graph
is clearly distinct for two regions: 
hard chaos ($1<w<2$) and soft chaos ($0<w<1$). But for larger $\Delta$,
the distinction between two regions starts to disappear.
}
\end{figure}

In Fig.\ \ref{widths}, we calculate $\bar{d_f}$ as the width $w$ varies from 0.5
to 2 with a step size of 0.02 
for seven opening sizes
($\Delta=1^\circ$, $\Delta=5^\circ$, $\Delta=10^\circ$, 
$\Delta=20^\circ$, $\Delta=30^\circ$, $\Delta=40^\circ$, and
$\Delta=50^\circ$), using the 2D set.
We can compare graphs in two different regions (found in 
the closed cut-circle billiard): $0<w<1$ (soft chaos)
and $1<w<2$  (hard chaos).
We observed similar behaviors seen in results for the 1D set in Ref.\
\onlinecite{ree7}.
However $\bar{d_f}$ does not reach one 
($\bar{d_f}=1$ represents \emph{non-chaos}), 
even when the billiard is integrable ($w=1$ or 2),
and this is due to the fine structure near trajectories 
bouncing very close to the circular boundary.
When the opening size $\Delta$ is not big (see graphs for 
$\Delta=5^\circ$, $\Delta=10^\circ$, and $\Delta=20^\circ$),
we observe that the behavior in two regions is clearly distinct.
In the region $0<w<1$, there are fluctuations, which comes from 
the mixed phase space structures
of the billiard, 
and in the region $1<w<2$,
graphs are smooth because the phase spaces of the billiard 
have no structure due to ergodicity.
On the other hand, when the opening size gets bigger 
(see graphs for $\Delta=30^\circ$,
$\Delta=40^\circ$, and $\Delta=50^\circ$), the  distinction between
two regions, observed in cases with smaller openings,  disappears.
There are fluctuations in both regions. 

In this Letter, we calculated the fractal dimension for the cut-circle
billiard with two openings by introducing the simplified box-counting algorithm 
for the 2D set of initial conditions.
The simplified box-counting algorithm introduced here is possible
because the function used to find singularities is constant
except singular points (i.\ e., the function is constant with
different values in all regions surrounded by singular boundaries).
With this method, some boxes containing singular points will be missed,
but the number of missed boxes is negligible
when we calculate the fractal dimension with big enough $N_P$ values.
Results for the 2D set were close to those for the 1D set 
only when $\Delta$ is small enough;
hence using the 1D set in fractal analysis does not 
fully represent the chaoticity of the billiard
for most cases.
In conclusion, this kind of fractal analyses gives us one of the fundamental
measures of the chaoticity of classical billiard systems,
and finding the relations with the quantum and semiclassical dynamics
of the same kind of billiards will be one of the interesting future works.

\begin{acknowledgments}
This work was supported by Kongju National University.
\end{acknowledgments}

\bibliography{p8}

\end{document}